\documentclass [12pt]{article}

\begin{document}
\begin{center}
\textbf{Wheeler-Feynman Absorber Theory Viewed by Model}
\end{center}

\begin{center}
\textbf{of Expansive Nondecelerative Universe} 
\end{center}

\bigskip

\begin{center}
Miroslav S\'uken\'{\i}k and Jozef \v{S}ima
\end{center}

\begin{center}
Slovak Technical University, Radlinsk\'eho 9, 812 37 Bratislava, Slovakia
\end{center}

\bigskip

\textbf{Abstract}. The present contribution documents the harmony of 
postulates and conclusions of Wheeler-Feynman absorber theory and the model 
of Expansive Nondecelerative Universe. A relationship connecting advanced 
electromagnetic waves and gravitational field quanta is rationalized.

\subsection*{Advanced and retarded waves in absorber theory}

The absorber theory was developed by Wheeler and Feynman in the forties [1] 
and elaborated by many others [2]. One of the features of modern physics is 
a renaissance of the above theory, its postulates, predictions, and 
consequences. The theory was aimed at the rationalization of radiation 
resistance. 

During its accelerated motion a charged particle -- a source of radiation - 
emits electromagnetic waves. These waves, known as classic retarded waves, 
are absorbed by another charged particle -- absorber - that, in turn, 
produces both retarded and advanced waves, the latter being characterized by 
negative energy. Advanced waves move backward in time to the emitter 
(instantaneous effect) and may act as a bearer of information. 

Evaluating the problem in more detail from the viewpoint of mathematics, 
differential equation describing the propagation of an emitted wave is 
second order in both space and time and provides thus two independent time 
solutions and two independent space solutions. It follows from a 
time-symmetrical solution of Maxwell equations [2]

\begin{equation}
\label{eq1}
\nabla \times E_{\pm}  = - {\frac{{dB_{\pm} } }{{dt}}}
\end{equation}

\noindent
that as to the vector of electric field intensity $E_{ +}  $, negative 
energy is attributed to the advanced waves, and as to the vector $E_{ -}  $, 
positive energy relates to the retarded waves. Denoting the electric field 
vector $E$ (or the magnetic field vector $B$) of the wave as $F$, then the 
retarded and advanced waves can be defined by the following equations [2]

\begin{equation}
\label{eq2}
F_{ret} = - i.F_{0} .\exp \left[  i(k.r - \omega .t)  \right]
\end{equation}

\begin{equation}
\label{eq3}
F_{adv} = i.F_{0} .\exp  \left[ i( - k.r + \omega .t)  \right]
\end{equation}

\noindent
where $k$ is the propagation vector, $r$ is the distance from the source of 
radiation, $\varpi $ is the frequency of the wave, and $t$ is the time. It 
follows from (\ref{eq2}) and (\ref{eq3}) that the retarded wave represents the 
positive-energy solution and reaches a point at the distance $r$ after the 
instant of radiation emission. Contrary, the advanced wave is the 
negative-energy (and negative frequency) solution and comes to a point at 
the distance $r$ before the instant of radiation emission.

Wheeler-Feynman absorber theory can be understood as an alternative to the 
classic probability interpretation of the Copenhagen school. It can 
contribute to explanation of several known paradoxes, such as that of 
Schr\"{o}dinger cat, quantum mechanical Einstein-Podolsky-Rosen locality 
premise, experiment with two slits, etc. [3, 4].

\subsection*{Gravitation in Expansive Nondecelerative Universe}

In the Expansive Nondecelerative Universe (ENU) model [5-7], negative values 
are attributed to the quanta of gravitational field, i.e. such quanta might 
represent a good candidate for advanced waves. Gravitational output $P_{g} $ 
of a body with the mass $m$is defined in the ENU as follows

\begin{equation}
\label{eq4}
P_{g} = - {\frac{{d}}{{dt}}}\int {{\frac{{R.c^{4}}}{{8\pi .G}}}dV = - 
{\frac{{m.c^{3}}}{{a}}}} 
\end{equation}

\noindent
where $a$ is the gauge factor of the Universe (its present calculated value 
is $1.3\times 10^{26}$ m) and $R$ is the scalar curvature (contrary to a 
more frequently used Schwarzschild metric in which $R = 0,$ in Vaidya metric 
applied in the ENU, $R \ne 0$ also outside a body).

The wave function of gravitational quanta is in the ENU described as

\begin{equation}
\label{eq5}
\Psi _{g} = \exp \left[  i.t.\left( {{\frac{{m.c^{5}}}{{\hbar .a.r^{2}}}}} 
\right)^{1 / 4}
 \right]
\end{equation}

In order to preserve the consistency of our postulates, it must hold

\begin{equation}
\label{eq6}
\alpha .P_{e} = {\left| {P_{g}}  \right|}
\end{equation}

\noindent
where $\alpha $ is the fine structure constant and $P_{e} $ is the 
electromagnetic (radiative) output of an accelerated charged particle. It 
follows from (\ref{eq4}), (\ref{eq5}), and (\ref{eq6}) that

\begin{equation}
\label{eq7}
\Psi _{g} = \exp \left[ i.t.\left( {{\frac{{\alpha .P_{e} .c^{2}}}{{\hbar 
.r^{2}}}}} \right)^{1 / 4}  \right]
\end{equation}

\subsection*{Gravitational quanta and advanced waves}

Postulating that the emitter must interact with more than one charged 
particles, the mass $m$ in (\ref{eq4}) represents a total mass of all charged 
particles interacting with the emitter during its accelerated orbital 
motion. In (\ref{eq5}) and (\ref{eq7}), $r$ is the mean distance of the emitter from the 
above mass of absorbers and, at the same time, it represents a mean radius 
of the emitter orbit curvature at its motion. Further, a 
Schr\"{o}dinger-type relation must hold

\begin{equation}
\label{eq8}
i.\hbar .{\frac{{d\Psi _{g}} }{{dt}}} = E_{g} .\Psi _{g} 
\end{equation}

Stemming from (\ref{eq7}) and (\ref{eq8}) it follows that

\begin{equation}
\label{eq9}
{\left| {E_{g}}  \right|} = \left( {{\frac{{\alpha .P_{e} .c^{2}.\hbar 
^{3}}}{{r^{2}}}}} \right)^{1 / 4}
\end{equation}

\noindent
where the radiation output $P_{e} $ of a charged emitter moving with the 
acceleration $g$ is formulated as

\begin{equation}
\label{eq10}
P_{e} = {\frac{{e^{2}.g^{2}}}{{6\pi .\varepsilon _{o} .c^{3}}}} \cong 
{\frac{{\alpha .\hbar .g^{2}}}{{c^{2}}}}
\end{equation}

Based on (\ref{eq9}) and (\ref{eq10}) it holds

\begin{equation}
\label{eq11}
{\left| {E_{g}}  \right|} \cong \hbar \left( {{\frac{{\alpha .g}}{{r}}}} 
\right)^{1 / 2}
\end{equation}

Simultaneously, $E_{g} $ must be identical to the energy $E_{adv} $ of 
advanced waves

\begin{equation}
\label{eq12}
E_{g} = E_{adv} 
\end{equation}

For synchrotronic radiation the rate $v$ of particles approach the rate of 
light $c$

\begin{equation}
\label{eq13}
v \cong c
\end{equation}

\noindent
and acceleration is nearly

\begin{equation}
\label{eq14}
g \cong {\frac{{c^{2}}}{{r}}}
\end{equation}

In such a limiting case the advanced and retarded waves can be described by 
functions

\begin{equation}
\label{eq15}
\Psi _{adv} = \exp\left[  i.t.\left( {{\frac{{\alpha .g}}{{r}}}}
  \right)^{1 / 2}
 \right]
\end{equation}

\begin{equation}
\label{eq16}
\Psi _{ret} = \exp \left[ - i.t.\left( {{\frac{{\alpha ^{1 / 2}.g}}{{c}}}} 
\right)
 \right]
\end{equation}

Using (\ref{eq13}) and (\ref{eq14}) the following equality can be evidenced

\begin{equation}
\label{eq17}
E_{g} = E_{adv} = - E_{ret} 
\end{equation}

\subsection*{Conclusion}

A target of the present contribution does not lie in evaluation of the 
correctness of Wheeler-Feynman absorber theory. Based on the validity of a 
logic assumption (\ref{eq6}), it is rather aimed at offering gravitational field 
quanta as a suitable candidate of advanced waves.

\subsection*{References}
\begin{enumerate}

\item
J.A. Wheeler, R.P. Feynman, Rev. Mod. Phys., 17 (1945) 157; 21 (1949) 425

\item
J.G. Cramer, Phys. Rev. D, 22 (1980) 362

\item
A. Einstein, B. Podolsky, N. Rosen, Phys. Rev., 47 (1935) 777

\item
J. Gribbin, Schrodinger's Kittsen, Weinedfeld \& Nicolson, London, 1995 

\item
V. Skalsk\'y, M. S\'uken\'{\i}k, Astrophys. Space Sci., 178 (1991) 169, 

\item
M. S\'uken\'{\i}k, J. \v{S}ima, J. Vanko, gr-qc/0010061

\item
M. S\'uken\'{\i}k, J. \v{S}ima, Astrophys. Space Sci., submitted

\end{enumerate}

\end{document}